\documentclass[proof]{WileyASNA-v1}

\usepackage{graphicx}

\newcommand{\ion}[2]{#1\,{\sc #2}}
\newcommand\farcs{\mbox{$.\!\!^{\prime\prime}$}}% 
\newcommand\arcsec{\mbox{$^{\prime\prime}$}}%

\articletype{Article Type}%

\received{26 April 2016}
\revised{6 June 2016}
\accepted{6 June 2016}

\begin{document}

\title{
Detection of a centrifugal magnetosphere in one of the most massive stars in 
the $\rho$\,Oph star-forming cloud
}

\author[1]{S. Hubrig*}
\author[2]{M. Sch\"oller}
\author[1]{S. P. J\"arvinen}
\author[1]{M. K\"uker}
\author[3]{A. F. Kholtygin}
\author[4]{P. Steinbrunner}

\authormark{S.~Hubrig \textsc{et al}}

\address[1]{\orgname{Leibniz-Institut f\"ur Astrophysik Potsdam (AIP)}, \orgaddress{An der Sternwarte~16, 14482~Potsdam, \country{Germany}}}
\address[2]{\orgname{European Southern Observatory}, \orgaddress{Karl-Schwarzschild-Str.~2, 85748~Garching, \country{Germany}}}
\address[3]{\orgdiv{Astronomical Institute}, \orgname{Saint-Petersburg State University}, \orgaddress{Universitetskij pr.~28, 198504 Saint-Petersburg, \country{Russia}}}
\address[4]{\orgname{Freie Universit\"at Berlin}, \orgaddress{Kaiserswerther Str.~16-18, 14195~Berlin, \country{Germany}}}

\corres{*Swetlana Hubrig. \email{shubrig@aip.de}}

\abstract{
Recent XMM-Newton observations of the B2 type star $\rho$\,Oph\,A indicated a periodicity of 1.205\,d,
which was ascribed to rotational modulation.
Since variability of X-ray emission in massive stars 
is frequently the signature of a magnetic field, we investigated whether
the presence of a magnetic field can indeed be invoked to explain the observed X-ray peculiarity. 
Two FORS\,2 spectropolarimetric observations in different rotation phases
revealed the presence of a negative ($\left< B_{\rm z}\right>_{\rm all}=-419\pm101$\,G) and
positive ($\left< B_{\rm z}\right>_{\rm all}=538\pm69$\,G) longitudinal magnetic field, respectively.
We estimate a lower limit for the dipole strength as $B_{\rm d} = 1.9\pm0.2$\,kG.
Our calculations of the Kepler and Alfv\'en radii imply the presence of a centrifugally
supported, magnetically confined plasma around $\rho$\,Oph\,A.
The study of the spectral variability indicates a behaviour similar to that observed in typical 
magnetic early-type Bp stars.
}

\keywords{
stars: early-type,
stars: individual:  $\rho$\,Oph\,A,
stars: magnetic field,
stars: variables: general,
stars: formation,
stars: X-rays
}

\maketitle

\section{Introduction}
\label{sect:intr}

The $\rho$\,Ophiuchus star-forming cloud is one of the closest low to intermediate mass star-forming regions 
and is known to contain the star-forming cluster LDN\,1688. 
Due to its youth, the relative 
proximity (120--160\,pc; e.g.\ \citeauthor{motte1998}, \citeyear{motte1998}), and its richness in young stars and protostars, 
the $\rho$\,Oph star-forming cloud has been actively investigated in recent years, using multiwavelength 
observations from X-ray to radio bands.

The radiation field in LDN\,1688 is dominated by two B-type stars, $\rho$\,Oph\,A (HD\,147933) 
and $\rho$\,Oph\,B (HD\,147934), 
both of spectral type B2V \citep{Abergel1996}, making up the visual
binary system $\rho$\,Oph\,AB. The apparent distance between 
the two stars is about 3$\farcs$1, and their orbital period is around 2400\,yr.
$\rho$\,Oph\,AB is surrounded by the bright and extended blue reflection nebula vdB\,105.
The basic structure of the cloud complex and its surroundings is described in detail by 
\citet{wilking2008}.

Massive young stars are known to emit strong X-rays. Unlike the X-ray emission from lower mass stars, 
which arises in stellar photospheres, the X-rays from massive stars are thought to result 
from powerful shocks. Detection of hard X-ray emission in massive stars 
appears frequently to be a signature of the presence of strong magnetic fields 
(e.g.\ \citeauthor{skin2008}, \citeyear{skin2008}).
The first XMM-Newton observations of $\rho$\,Oph\,AB over 53\,ks were obtained by
\citet{pil2014}.
Their analysis showed a smooth variability of the X-ray emission, probably caused 
by the emergence of an extended active region on the surface of $\rho$\,Oph\,A. 
The derived hardness ratios were periodic, with the hardest spectrum corresponding to the
highest count rate.
According to \citet{pil2014}, the observations are fully compatible with the hypothesis of
the presence of a region brighter and hotter than the average stellar surface that
gradually appears on the visible side of the star during the rise
of the count rate.
Follow-up XMM-Newton observations with a duration of 140\,ks in 2016
allowed \citet{pil2017} to detect a periodicity of 1.205\,d,
which they ascribed to rotational modulation of the X-ray emission.  Analysing the time resolved X-ray spectra,
the authors speculated that either intrinsic
magnetism produces a hot spot on the surface  of $\rho$\,Oph\,A, or an unknown low mass companion is 
the source of the observed X-ray variability.
Clearly, the remarkable behavior of $\rho$\,Oph\,A in X-rays deserves further investigations to find out
whether the presence of a magnetic field can indeed be invoked to explain the observed X-ray peculiarity.

In Section~\ref{sect:data}, we give an overview about the FORS\,2 spectropolarimetric observations of $\rho$\,Oph\,A
as well as the data reduction,
and describe the results of the magnetic field measurements.
The magnetospheric parameters and
our analysis of the spectral variability of $\rho$\,Oph\,A
are shown in Sections~\ref{sect:mag_param} and \ref{sect:var}. 
A discussion of our results is
presented in Section~\ref{sect:disc}.

\section{Observations and magnetic field measurements}
\label{sect:data}

\begin{table*}
\caption{
Logbook of the FORS\,2 spectropolarimetric observations of $\rho$\,Oph\,A,
including the modified Julian date of mid-exposure,
followed by the achieved signal-to-noise ratio in the Stokes~$I$ spectra around 5200\,\AA{},
and the measurements of the mean longitudinal magnetic field using the Monte Carlo bootstrapping test,
for all lines and for the hydrogen lines.
In the last columns,
we present the results of our measurements using the null spectra for the set
of all lines and the orbital phases.
The rotation phases of $\rho$\,Oph\,A are calculated relative
to a zero phase corresponding to the date of the first FORS\,2 observation at 
MJD\,57951.2242 assuming $P_{\rm rot}=1.205$\,d.
All quoted errors are 1$\sigma$ uncertainties. 
}
\label{tab:log_meas}
\centering
\begin{tabular}{lrr@{$\pm$}rr@{$\pm$}rr@{$\pm$}rl}
\hline
\hline
\multicolumn{1}{c}{MJD} &
\multicolumn{1}{c}{SNR} &
\multicolumn{2}{c}{$\left< B_{\rm z}\right>_{\rm all}$} &
\multicolumn{2}{c}{$\left< B_{\rm z}\right>_{\rm hyd}$} &
\multicolumn{2}{c}{$\left< B_{\rm z}\right>_{\rm N}$} &
\multicolumn{1}{c}{$\varphi_{\rm orb}$}\\
&
\multicolumn{1}{c}{$\lambda$5200} &
\multicolumn{2}{c}{[G]} &
\multicolumn{2}{c}{[G]}  &
\multicolumn{2}{c}{[G]} &
\\
\hline
   57951.2242& 2297 & $-$419 & 101 &$-$301  &   142 & $-$72&   92& 0\\
   57976.0708& 3005 &    538 &  69 &   569  &    94 &    29 &  68& 0.620\\
\hline
\end{tabular}
\end{table*}

FOcal Reducer and low dispersion Spectrograph (FORS\,2; \citeauthor{Appenzeller1998}, \citeyear{Appenzeller1998}) 
spectropolarimetric observations of $\rho$\,Oph\,A
were obtained on 2017 July~17 and August 11.
The FORS\,2 multi-mode instrument is equipped with polarisation analysing optics
comprising super-achromatic half-wave and quarter-wave phase retarder plates,
and a Wollaston prism with a beam divergence of 22$\arcsec$ in standard
resolution mode. 
We used the GRISM 600B and the narrowest available slit width
of 0$\farcs$4 to obtain a spectral resolving power of $R\approx2000$.
The observed spectral range from 3250 to 6215\,\AA{} includes all Balmer lines,
apart from H$\alpha$, and numerous helium lines.
For the observations, we used a non-standard readout mode with low 
gain (200kHz,1$\times$1,low), which provides a broader dynamic range, hence 
allowed us to reach a higher signal-to-noise ratio (SNR) in the individual spectra.

A first description of the assessment of longitudinal magnetic field
measurements using FORS\,1/2 spectropolarimetric observations was presented 
in our previous work (e.g.\ \citeauthor{Hubrig2004a}, \citeyear{Hubrig2004a}, \citeyear{Hubrig2004b}, 
and references therein).
To minimize the cross-talk effect,
and to cancel errors from 
different transmission properties of the two polarised beams,
a sequence of subexposures at the retarder
position angles
$-$45$^{\circ}$$+$45$^{\circ}$,
$+$45$^{\circ}$$-$45$^{\circ}$,
$-$45$^{\circ}$$+$45$^{\circ}$,
etc.\ is usually executed during the observations. 
Moreover, the reversal of the quarter wave 
plate compensates for errors in the relative wavelength calibrations of the two
polarised spectra.
According to the FORS User Manual, the $V/I$ spectrum is calculated using:

\begin{equation}
\frac{V}{I} = \frac{1}{2} \left\{ 
\left( \frac{f^{\rm o} - f^{\rm e}}{f^{\rm o} + f^{\rm e}} \right)_{-45^{\circ}} -
\left( \frac{f^{\rm o} - f^{\rm e}}{f^{\rm o} + f^{\rm e}} \right)_{+45^{\circ}} \right\},
\end{equation}

\noindent
where $+45^{\circ}$ and $-45^{\circ}$ indicate the position angle of the
retarder waveplate and $f^{\rm o}$ and $f^{\rm e}$ are the ordinary and
the extraordinary beam, respectively. 
Null profiles, $N$, are calculated as pairwise differences from all available 
$V$ profiles.  From these, 3$\sigma$-outliers are identified and used to clip 
the $V$ profiles.  This removes spurious signals, which mostly come from cosmic
rays, and also reduces the noise. A full description of the updated data 
reduction and analysis will be presented in a separate paper (Sch\"oller et 
al., in preparation, see also \citeauthor{Hubrig2014a}, \citeyear{Hubrig2014a}).
The mean longitudinal magnetic field, $\left< B_{\rm z}\right>$, is 
measured on the rectified and clipped spectra based on the relation 
following the method suggested by \citet{angel1970}

\begin{eqnarray} 
\frac{V}{I} = -\frac{g_{\rm eff}\, e \,\lambda^2}{4\pi\,m_{\rm e}\,c^2}\,
\frac{1}{I}\,\frac{{\rm d}I}{{\rm d}\lambda} \left<B_{\rm z}\right>\, ,
\label{eqn:vi}
\end{eqnarray} 

\noindent 
where $V$ is the Stokes parameter that measures the circular polarization, $I$
is the intensity in the unpolarized spectrum, $g_{\rm eff}$ is the effective
Land\'e factor, $e$ is the electron charge, $\lambda$ is the wavelength,
$m_{\rm e}$ is the electron mass, $c$ is the speed of light, 
${{\rm d}I/{\rm d}\lambda}$ is the wavelength derivative of Stokes~$I$, and 
$\left<B_{\rm z}\right>$ is the mean longitudinal (line-of-sight) magnetic field.

The longitudinal magnetic field was measured in two ways: using the entire spectrum
including all available lines
or using exclusively the hydrogen lines.
Furthermore, we have carried out Monte Carlo bootstrapping tests. 
These are most often applied with the purpose of deriving robust estimates of standard errors. 
The measurement uncertainties obtained before and after the Monte Carlo bootstrapping tests were found to be 
in close agreement, indicating the absence of reduction flaws. 
The results of our magnetic field measurements, those for the entire spectrum
or only for the hydrogen lines, are presented in 
Table~\ref{tab:log_meas}. As no ephemeris is known for $\rho$\,Oph\,A -- only $P_{\rm rot}=1.205$\,d
is mentioned in \citet{pil2017} -- we fixed the rotation phase 0 at the date of the 
first FORS\,2 observation at MJD\,57951.2242.

The magnetic field of $\rho$\,Oph\,A was detected on both observing dates. Using the entire spectrum,
we measure a field of negative polarity, $\left< B_{\rm z}\right>_{\rm all}=-419\pm101$\,G, 
at a significance level
of 4.1$\sigma$ in the data obtained on 2017 July 17, while the measurement using the hydrogen lines 
shows a significance of 2.1$\sigma$.
The highest values for the longitudinal
magnetic field, $\left< B_{\rm z}\right>_{\rm all}=538\pm69$\,G at a significance level
of 7.8$\sigma$ using the entire spectrum and $\left< B_{\rm z}\right>_{\rm hyd}=569\pm94$\,G at a significance level
of 6.1$\sigma$ using the hydrogen lines was measured in the data obtained on 2017 August 11.
No detection was achieved in the diagnostic $N$ spectra,
indicating the absence of spurious polarization signatures.

\begin{figure}
\centering
\includegraphics[width=0.23\textwidth]{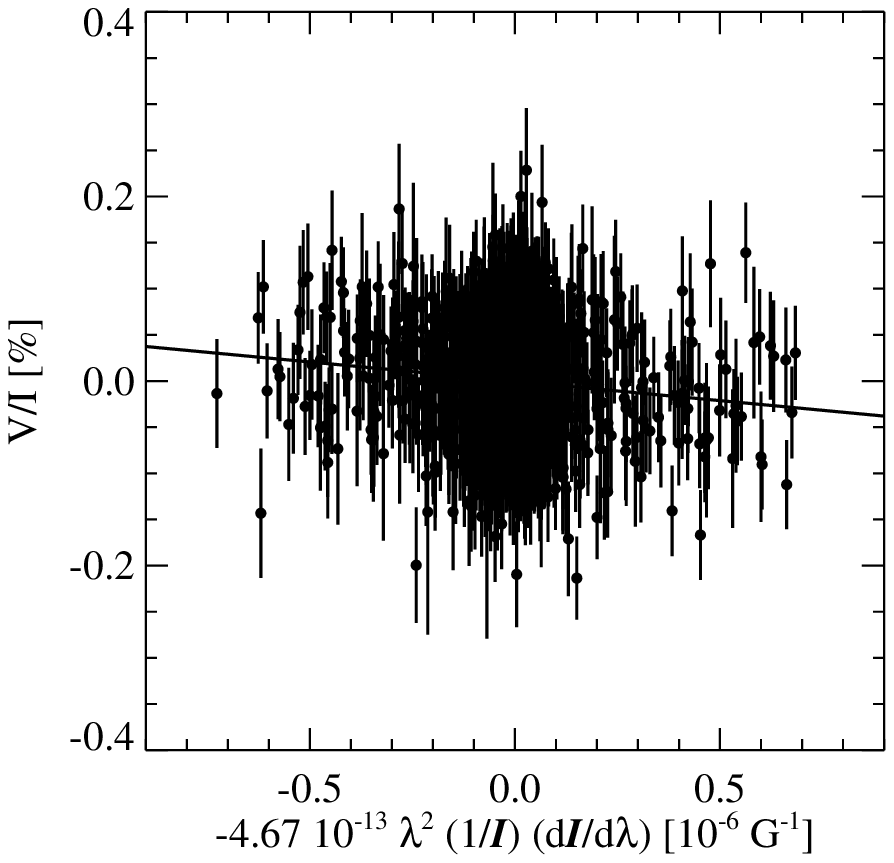}
\includegraphics[width=0.23\textwidth]{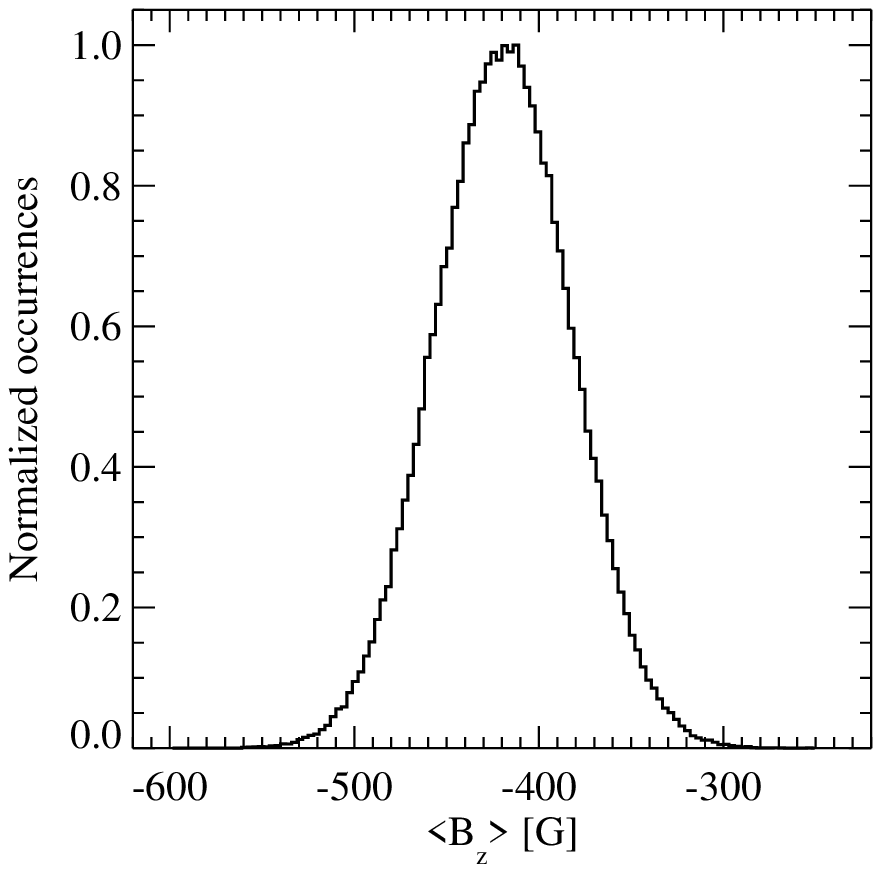}
\caption{
{\it Left panel}: Linear fit to Stokes~$V$
obtained for the FORS\,2 observation of $\rho$\,Oph\,A on MJD\,57951.2242.
{\it Right panel}: Distribution of the longitudinal magnetic field values $P(\left<B_{\rm z}\right>)$, 
which were obtained 
via bootstrapping. From this distribution follows the most likely 
value for the longitudinal 
magnetic field $\left< B_{\rm z}\right>_{\rm all}=-419\pm101$\,G.
}
\label{fig:rho1}
\end{figure}

\begin{figure}
\centering
\includegraphics[width=0.23\textwidth]{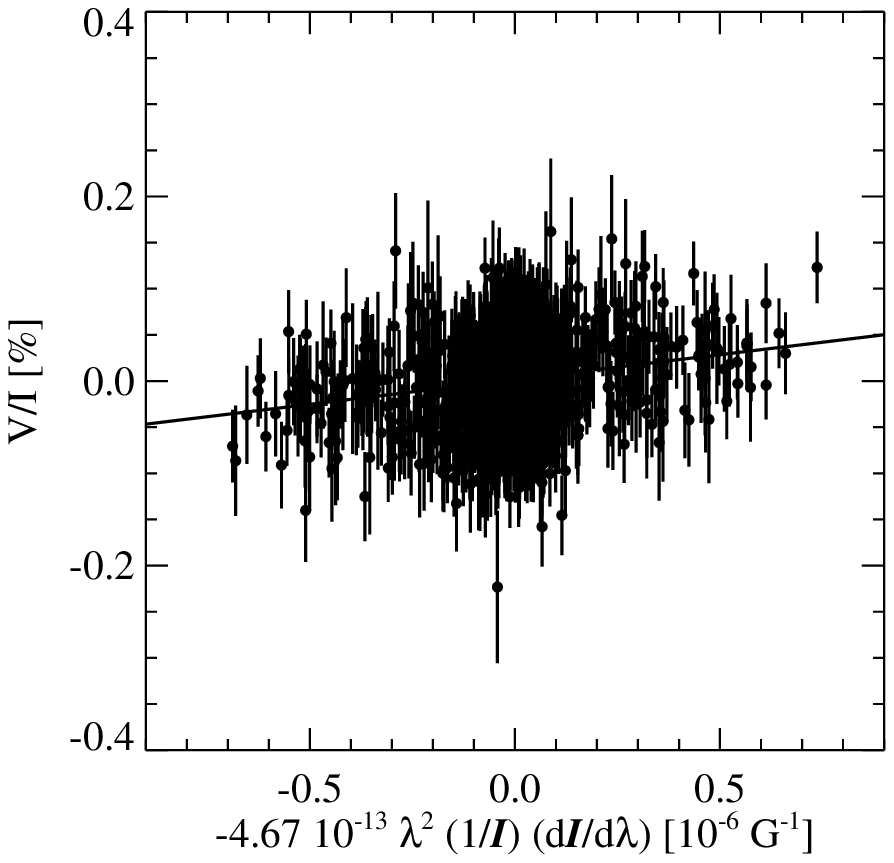}
\includegraphics[width=0.23\textwidth]{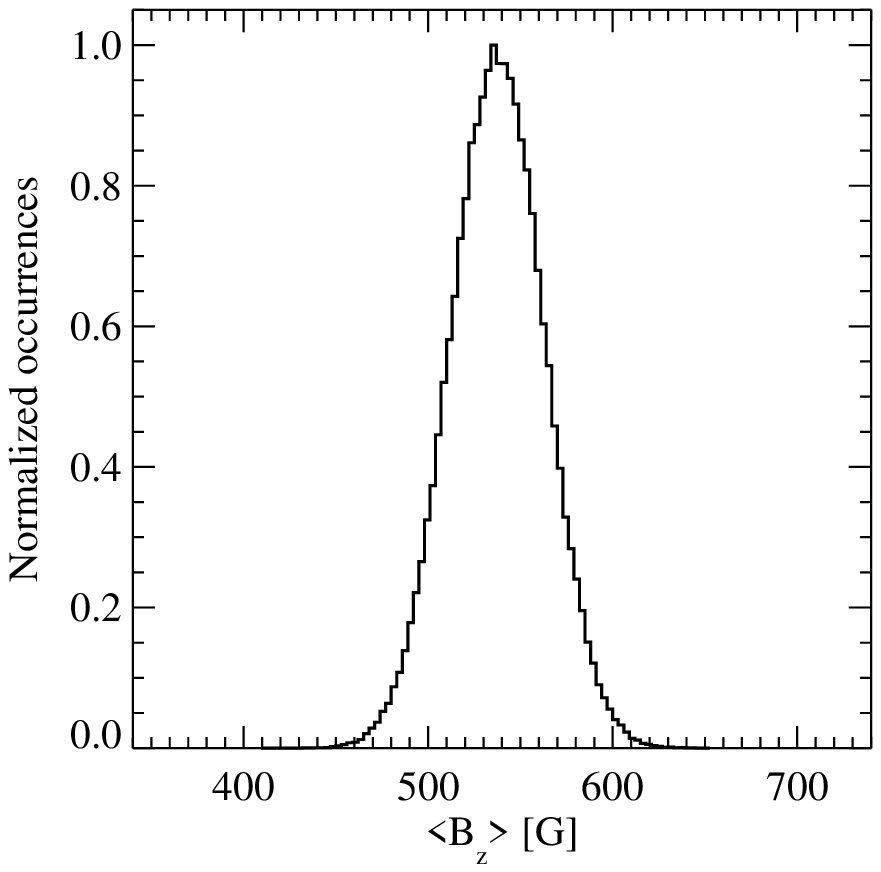}
\caption{
{\it Left panel}: Linear fit to Stokes~$V$
obtained for the FORS\,2 observation of $\rho$\,Oph\,A on MJD\,57976.0708.
{\it Right panel}: Distribution of the longitudinal magnetic field values $P(\left<B_{\rm z}\right>)$, 
which were obtained 
via bootstrapping. From this distribution follows the most likely 
value for the longitudinal 
magnetic field $\left< B_{\rm z}\right>_{\rm all}=538\pm69$\,G.
}
\label{fig:rho2}
\end{figure}

\begin{figure}
\centering
\includegraphics[width=0.45\textwidth]{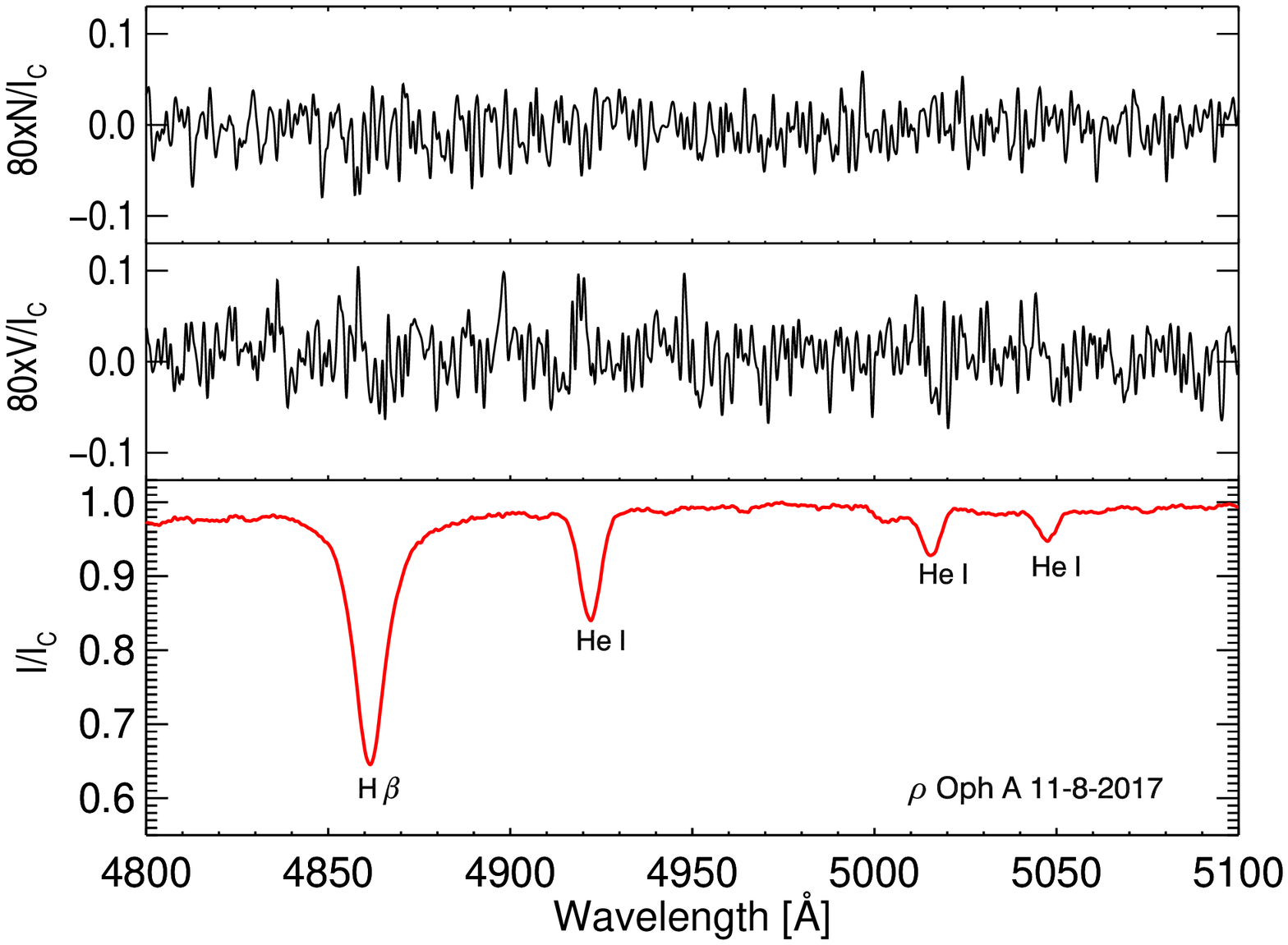}
\caption{ 
Stokes~$I$, $V$, and diagnostic $N$ spectra (from bottom to top) of $\rho$\,Oph\,A
in the vicinity of the H$\beta$ line.
Note that the Stokes~$V$ and the diagnostic $N$ spectra were magnified by a factor of 80.
}
\label{fig:zf}
\end{figure}

In Figs.~\ref{fig:rho1} and \ref{fig:rho2}, we show the linear regressions in plots of
$V/I$ against $-4.67\,10^{-13} \lambda^2 (1/I) ({\rm d}I/{\rm d}\lambda)$ together with 
the results of the Monte Carlo bootstrapping tests.
In Fig.~\ref{fig:zf}, we present Stokes $I$, $V$, and diagnostic $N$ spectra of $\rho$\,Oph\,A obtained on 2017 
August 11 in the spectral region around the H$\beta$ line and several \ion{He}{i} lines. 

We can estimate the dipole strength of $\rho$\,Oph\,A  following 
the model by \citet{stibbs1950} as formulated by \citet{preston1967}:

\begin{eqnarray}
B_{\rm d} \ge & \left< B_{\rm z}\right>^{\rm max}  \left( \frac{15+u}{20(3-u)}\right)^{-1}.
\end{eqnarray}

Assuming a limb-darkening coefficient $u$ of 0.3, typical for the spectral type B2V \citep{claret2011},
we can give a lower limit for the dipole strength of $B_{\rm d} \ge 1.9\pm0.2$\,kG.

\section{Magnetospheric parameters}
\label{sect:mag_param}

Similar to the small number of previously studied early-B type stars, 
the X-ray emission in $\rho$\,Oph\,A detected in XMM-Newton observations can be generated via magnetically 
confined shocks.
\citet{babel1997} suggested that in stars with large-scale magnetic fields 
wind streams from opposite hemispheres are channeled
toward the magnetic equator, where they collide, leading to strong shocks and associated X-rays.
To investigate whether the wind plasma is locked to the magnetic field,
we need to know  several
physical parameters of $\rho$\,Oph\,A.
Currently, the only information
can be found in the work of \citet{pil2017}, who 
assumed a stellar radius of $\sim8\,R_\odot$, without mentioning the method of estimation,
used the rotational 
modulation of the X-ray emission to determine a rotation period of 1.205\,d,
and applied a Fourier transform 
to the \ion{He}{i} 6678 line to measure $v$\,sin\,$i=239.5\pm10$\,km\,s$^{-1}$.

Based on revised trigonometric parallaxes from the Hipparcos data \citep{Leeuwen2008},
\citet{mamajek2008} concluded that 
a distance of $131\pm3$\,pc to the Ophiuchus star-forming region is the best available 
derived from Hipparcos data.
Using this distance, the extinction $A_{\rm v}=3$ from \citet{pil2016}, 
assuming $T_{\rm eff}=21\,000$\,K for the spectral type B2 \citep{bohm1981}, and 
the corresponding bolometric correction $BC=-2.0\pm0.1$ \citep{flow1996},
we estimate $\log \left({L}/{L_\odot}\right) = 4.1\pm0.2$, taking into account estimation inaccuracies 
of the distance determination and the bolometric correction.
From the position of $\rho$\,Oph\,A in the H-R diagram, using evolutionary tracks from
\citet{Ekstroem2012} and assuming that $\rho$\,Oph\,A is still in the hydrogen fusing stage,
we find a mass of about $10\pm0.7\,{M_\odot}$.
From the Stefan--Boltzmann law we find $\log(R/R_\odot) \approx 0.93$,
i.e.\ a value of $8.5\pm1\,{R_\odot}$ for the stellar radius. Using this radius, $v \sin i = 240\pm10$\,km\,s$^{-1}$,
and the rotation period $P_{\rm rot} = 1.205$\,d,  
we obtain $v_{\rm eq}=360\pm40$\,km\,s$^{-1}$ and an inclination angle of the stellar rotation axis to the line of
sight $i=42\pm6^{\circ}$.
Further, the rotation period of 1.205\,d
corresponds to an angular velocity $\Omega=6.035 \times 10^{-5} {\rm s}^{-1}$.
To determine the properties of the stellar wind and the magnetosphere,
we first compute the escape velocity,
$ v_{\rm esc} = \sqrt{{2 G M}/{R}} = 671$\,km\,s$^{-1}$.
Assuming that
$ {v_\infty}/{v_{\rm esc}} = 1.3$ \citep{Vink2001},
we find $v_\infty=873$\,km\,s$^{-1}$
for the terminal velocity.
Using equation~(25) of \citet{Vink2001},
we calculate a mass loss rate
$ \dot{M} = 8.69 \times 10^{-9} M_\odot/{\rm yr}$ for solar metallicity. 
With a polar magnetic field strength
$B_p \ge1.9$\,kG, we obtain a confinement parameter
$\eta_* = ({B_{\rm eq}^2 R_*^2})/({\dot{M} v_\infty}) \ge 6.55 \times 10^4$,
where $B_{\rm eq} = 0.5 B_p$ \citep{udDoula2002}.
The impact of rotation is measured by the parameter
$W={\Omega R} /{v_{\rm K}} = 0.75$,
where $v_K=\sqrt{G M / R}$. 
Using equation~(9) from \citet{udDoula2008}, we thus arrive at 
$ {R_A} = (0.3+ (\eta_*+0.25)^{1/4}) {R_*} \ge 9.3\,{R_*}$
for the Alfv\'en radius and
${R_K} = \left({ G M}/{\Omega^2}\right)^{1/3} = 1.21\,R_*$
for the Kepler (corotation) radius.

\citet{petit2013} divided magnetic massive stars into two
groups, those with dynamical magnetospheres (DMs) with ${R_A} < {R_K}$   and
those possessing centrifugal magnetospheres (CMs) with ${R_A} > {R_K}$. 
Since for $\rho$\,Oph\,A the radial extent of the magnetic confinement of the wind given by the Alfv\'en radius
is much larger than the Kepler radius, material caught in the region
between ${R_A}$ and ${R_K}$
is centrifugally supported against infall, and so builds up to a much denser CM. 

\begin{figure*}
\centering
\includegraphics[width=1.00\textwidth]{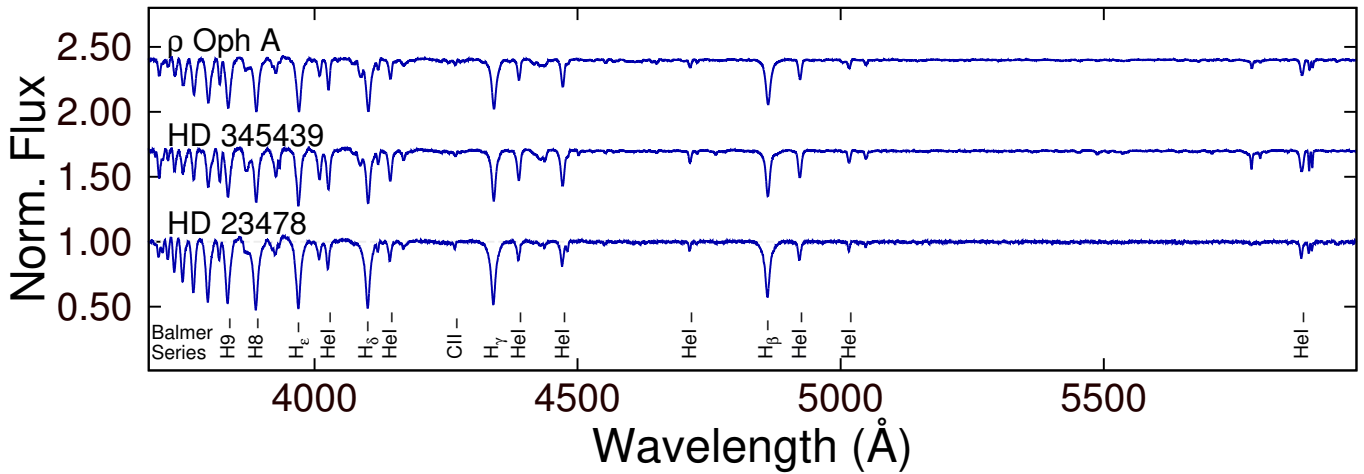}
\caption{ 
The normalized FORS\,2 Stokes~$I$
spectrum of $\rho$\,Oph\,A is displayed together with the normalized FORS\,2 spectra of two other rapidly 
rotating early-B type stars, 
HD\,23478 and HD\,345439, for which the presence of a rigidly rotating magnetosphere was recently detected 
\citep{Eikenberry2014}.
Well known spectral
lines are indicated. The spectra of HD\,345439 and  $\rho$\,Oph\,A were vertically offset for clarity.
}
\label{fig:norm}
\end{figure*}

The rotational modulation of the X-ray emission with $P_{\rm rot}=1.205$\,d detected by
\citet{pil2017} possibly indicates that 
the wind plasma is predominantly locked to the magnetic field and that the magnetosphere of $\rho$\,Oph\,A
can be interpreted in the context of the rigidly rotating magnetosphere model (RRM;  \citeauthor{Townsend2005}, \citeyear{Townsend2005}).
A few rapidly rotating early-B type stars with RRM magnetospheres showing comparable rotation periods and
magnetic field strengths were discovered in the last years (e.g., \citeauthor{riv2013}, \citeyear{riv2013},
\citeauthor{Eikenberry2014}, \citeyear{Eikenberry2014}).
As is shown in Fig.~\ref{fig:norm}, the  spectral  appearance of $\rho$\,Oph\,A in our FORS\,2 observations 
is very similar to the spectral appearance of 
two other rapidly rotating early-B type stars, 
HD\,23478 ($P_{\rm rot}=1.05$\,d) and HD\,345439 ($P_{\rm rot}=0.77$\,d),
for which the presence of RRMs was recently detected \citep{Eikenberry2014}.
In all stars with RRMs, the cooler and denser postshock material trapped in the stellar 
magnetospheres is typically detected in the H$\alpha$ line or in the near infrared hydrogen recombination lines.
Since our FORS\,2 polarimetric spectra of $\rho$\,Oph\,A do not cover the spectral region containing 
the H$\alpha$ line,
it would be valuable to monitor the variability of the H$\alpha$ line
profile in future observations to confirm that circumstellar gas is locked to the magnetosphere and is in corotation with the 
stellar surface.

\section{Spectral variability}
\label{sect:var}

\begin{figure*}
\centering
\includegraphics[width=0.45\textwidth]{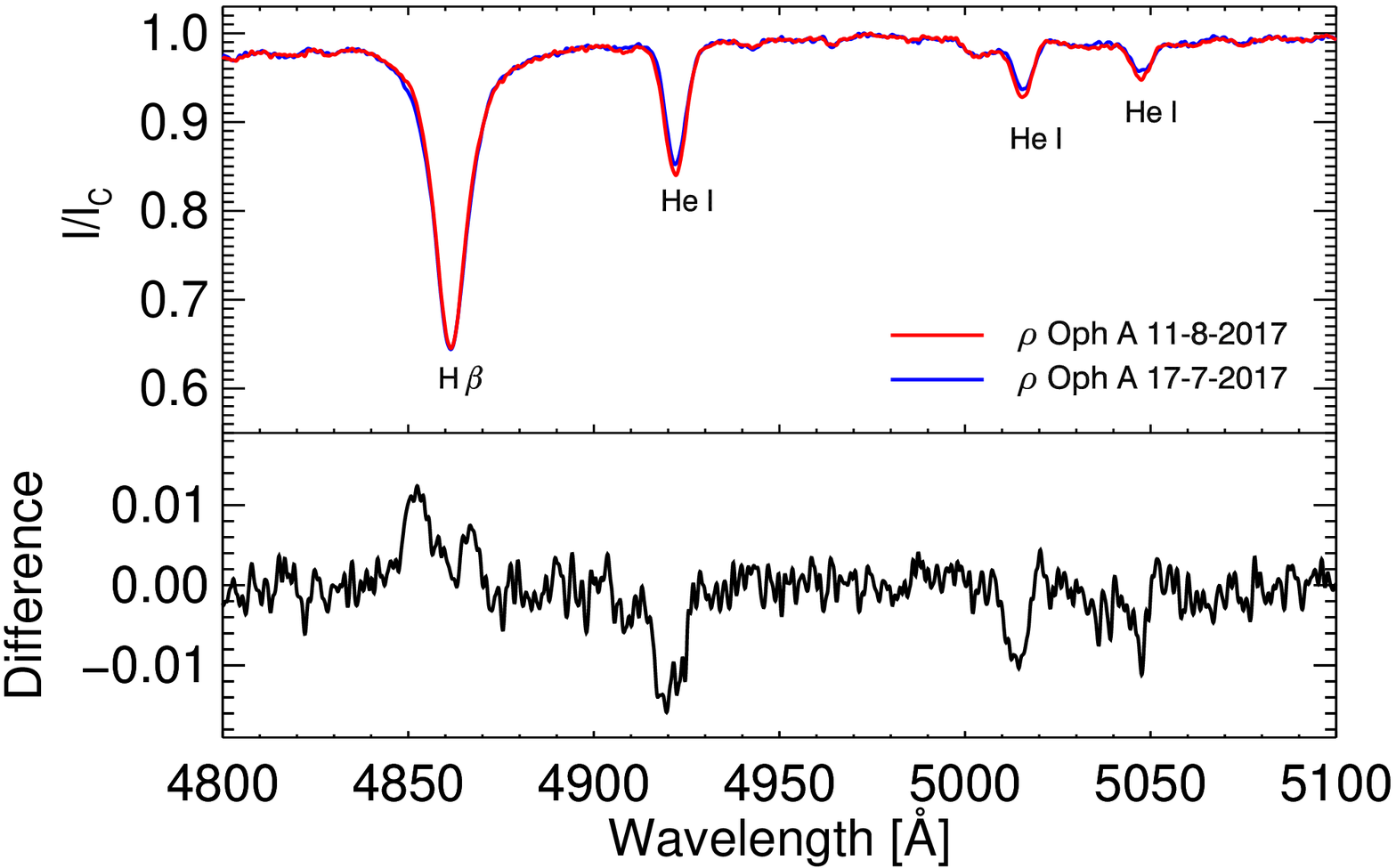}
\includegraphics[width=0.45\textwidth]{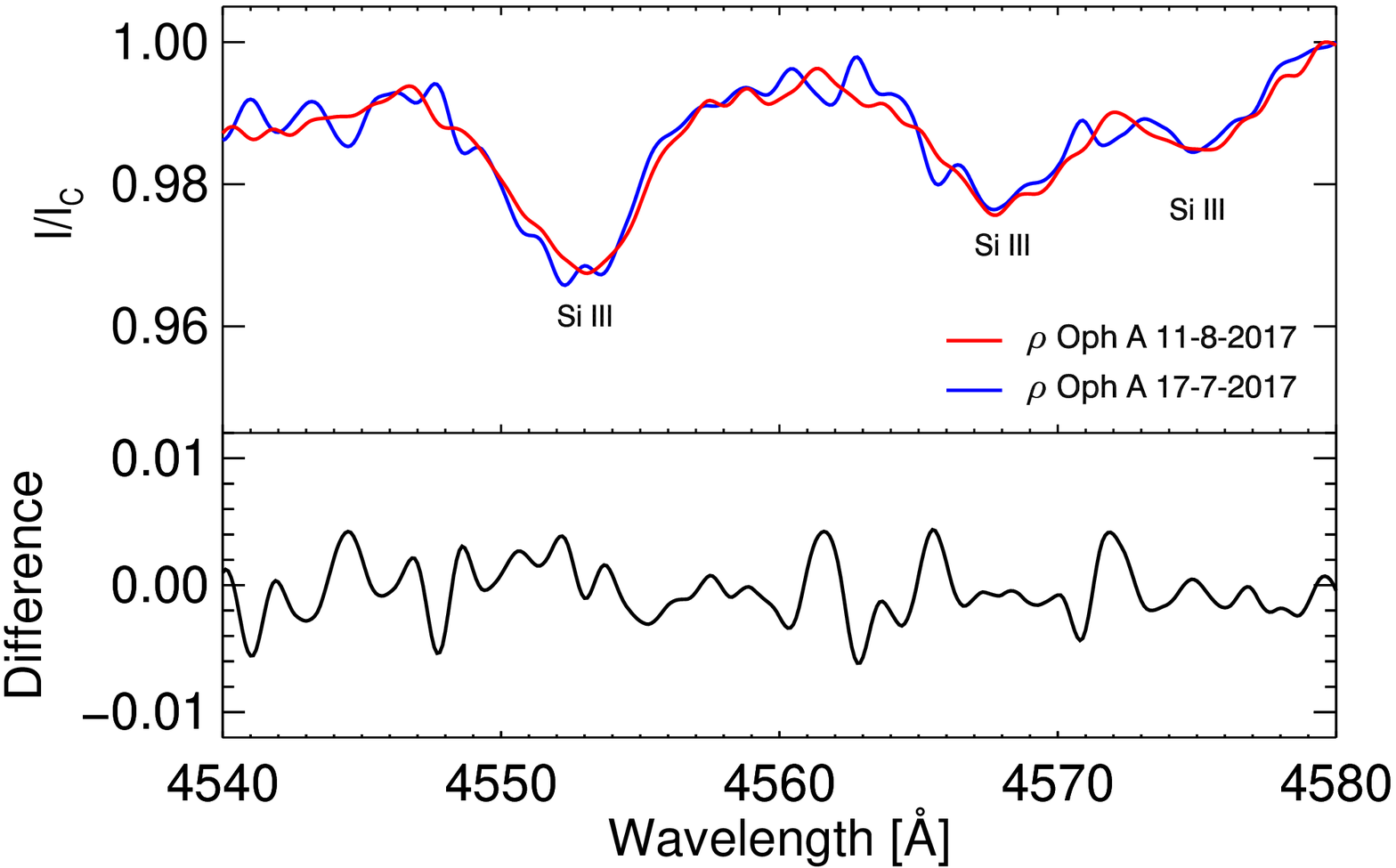}
\caption{
{\it Left panel}:
Variability of the H$\beta$ and \ion{He}{i} lines in the spectra of $\rho$\,Oph\,A. 
On the top one can see the normalized FORS\,2 spectra obtained on two different nights.
On the bottom, the difference between the two spectra is shown.
{\it Right panel}:
No variability is detected for the \ion{Si}{iii} lines at wavelengths 4553, 4567, and 4574\,\AA{}.
}
\label{fig:J11_1}
\end{figure*}

In early-B type magnetic Bp stars, the global magnetic dipole-like field is tilted to the rotation axis
by the angle $\beta$ and the surface distribution of
certain chemical elements, such as silicon or carbon, displays a spotted structure, which is, as a rule, 
disjunct from the helium distribution.  As the star rotates, we should detect variations in
the strength of the longitudinal magnetic field and the intensities of spectral line profiles of various elements
with the rotation period of the star. Since our observations correspond to two different rotational phases,
where opposite magnetic field polarities are detected, we have compared the 
line profile shapes on these two different epochs.  Due to the rather low FORS\,2 spectral resolution,
we checked the variability only for hydrogen, helium, and silicon lines. 
Our comparison of the line profiles belonging to these elements
presented in Fig.~\ref{fig:J11_1} reveals distinct changes in the line intensities of the helium lines, supporting
the assumption of the presence of an inhomogeneous helium distribution on the stellar surface. We also detect
that the H$\beta$ line intensity is lower in the phase where the helium line intensities are stronger.
On the other hand, silicon lines are very weak and noisy and do not present any obvious variability.
To check variability on a time scale of a few minutes, we compared the stability of 
the line profiles belonging
to these elements over the full sequences of sub-exposures obtained on that time-scale.
No short-term variability was detected in both FORS\,2 observations.

\section{Discussion}
\label{sect:disc}
 
The analysis of FORS\,2 spectropolarimetric observations of  $\rho$\,Oph\,A on two different rotation 
phases reveals the presence of a rather strong magnetic field with a dipole strength of $B_{\rm d} \ge 1.9\pm0.2$\,kG.
Using physical parameters of this B2 type star, we calculated Kepler and Alfv\'en radii and concluded
that a centrifugally
supported, magnetically confined plasma is present around $\rho$\,Oph\,A.
Since the magnetic field of $\rho$\,Oph\,A was measured only on two occasions, it should be a prime candidate 
for a follow-up spectropolarimetric study that would lead to more accurate magnetospheric parameters.

A comparison of line profiles on two different rotation phases shows a clear variability of helium lines
similar to that observed in typical magnetic early-type Bp stars. The variability of the H$\beta$ line
can probably be explained by the presence of an extended magnetosphere around $\rho$\,Oph\,A.
Based on our detection of the presence of a magnetic field in $\rho$\,Oph\,A, we conclude that the most 
likely reason for the variations of the X-ray emission observed by \citet{pil2017}
is the occultation of parts of the magnetosphere by the stellar body.

The origin of magnetic fields in massive stars
is still a major unresolved problem in astrophysics. Only a small fraction of stars
(5--7\%, e.g.\ \citeauthor{schol2017}, \citeyear{schol2017}) with
radiative envelopes possess strong large-scale organized magnetic fields. Such fields can 
probably be generated during the star formation process, by dynamo action taking
place in the rotating stellar cores, or they could be products of a
merger process. While the first two scenarios are unable to explain
a number of observational phenomena (e.g.\ \citeauthor{fer2015}, \citeyear{fer2015}),
 the magnetic fields might form when two protostellar objects merge late during their evolution 
towards the main sequence and when at least one of them has already acquired a radiative envelope
\citep{fer2009}.

$\rho$\,Oph\,A is not the only magnetic star detected in a complex star forming region.
In 2013, using the High Accuracy Radial velocity Planet Searcher polarimeter 
(HARPSpol; \citeauthor{snik2008}, \citeyear{snik2008})
attached to ESO's 3.6 m telescope (La Silla,
Chile) and FORS\,2 observations, \citet{Hubrig2014b} searched  for the 
presence of a magnetic field in the three most massive central stars in the Trifid nebula, HD\,164492A, 
HD\,164492C, and HD\,164492D. These observations indicated the presence of a strong longitudinal magnetic 
field of about 500--600\,G in the poorly studied component HD\,164492C.
Later, \citet{gonz2017} showed that HD\,164492C is a spectroscopic triple system 
consisting of an eccentric close spectroscopic binary with a period of 12.5\,d, and a massive 
fast rotating He-rich tertiary possessing a variable kG order magnetic field.
Similar to $\rho$\,Oph\,A, also in HD\,164492C the X-ray emission was firmly detected using Chandra
observations \citep{rho2004}.
The detection of magnetic massive stars
in the youngest star-forming regions implies
that these targets may play a pivotal role in our understanding of
the origin of magnetic fields in massive stars. 
It is striking that both magnetic massive stars located in young 
star-forming regions show similar characteristics such as fast rotation and the presence
of X-ray emission. Also their stellar surfaces show helium abundance variations typical for He-rich 
Bp stars with large-scale organized magnetic fields.
Especially intriguing is the presence of an extended blue reflection nebula around the system $\rho$\,Oph\,AB:
A few years ago, \citet{Hubrig2013a} discussed the variability of the 
longitudinal magnetic field in the O6.5f?p star HD\,148937 (\citeauthor{Hubrig2008}, \citeyear{Hubrig2008}, \citeyear{Hubrig2013b}),
suggesting that this target may provide a smoking gun, as it is 
surrounded by the 3000\,yr old, nitrogen-rich
bipolar nebula NGC\,6164/5 \citep{leith1987}, which was likely created 
through strong binary interaction. 
In parallel with HD\,148937, we can speculate that $\rho$\,Oph\,A can similarly be a merger product and that the
surrounding nebula is created by the ejected material. Obviously, it would be important to 
investigate the chemical composition of the material of the nebula around $\rho$\,Oph\,A to determine its origin.
Furthermore, since star formation in molecular clouds is assumed to be triggered by the dynamical action 
of winds from massive stars, we need to understand how magnetized winds from magnetic massive stars 
formed during the first episodes of star
formation influence their environments, including nearby sites of star formation and protoplanetary
disks surrounding low-mass pre-main-sequence stars.

\section*{Acknowledgments}
Based on observations made with ESO Telescopes at the La Silla Paranal Observatory
under programme 099.D-0067(A).
AK acknowledges financial support from RFBR grant 16-02-00604A.


\begin{thebibliography}{}

\bibitem[Abergel et al.(1996)]{Abergel1996}
Abergel, A., Bernard, J.~P., Boulanger, F., et al.\ 1996,
A\&A, 315, L329

\bibitem[Angel \& Landstreet(1970)]{angel1970}
Angel, J.~R.~P., Landstreet, J.~D.\ 1970,
ApJ, 160, L147

\bibitem[Appenzeller et al.(1998)]{Appenzeller1998}
Appenzeller, I., Fricke, K., F{\"u}rtig, W., et al.\ 1998,
The ESO Messenger, 94, 1

\bibitem[Babel \& Montmerle(1997)]{babel1997}
Babel, J., Montmerle, T.\ 1997,
A\&A, 323, 121

\bibitem[B\"ohm-Vitense(1981)]{bohm1981}
B\"ohm-Vitense, E.\ 1981,
Ann.\ Rev.\ Astron.\ Astrophys., 19, 295
%

\bibitem[Claret \& Bloemen(2011)]{claret2011}
Claret, A., Bloemen, S.\ 2011,
A\&A, 529, A75

\bibitem[Eikenberry et al.(2014)]{Eikenberry2014}
Eikenberry, S.~S., Chojnowski, S.~D., Wisniewski, J., et al.\ 2014,
ApJL, 784, L30

\bibitem[Ekstr\"om et al.(2012)]{Ekstroem2012}
Ekstr\"om, S., Georgy, C., Eggenberger, P., et al.\ 2012,
A\&A, 537, A146

\bibitem[Ferrario et al.(2009)]{fer2009}
Ferrario, L., Pringle, J.~E., Tout, C.~A., Wickramasinghe, D.~T.\ 2009,
MNRAS, 400, L71

\bibitem[Ferrario et al.(2015)]{fer2015}
Ferrario, L., Melatos, A., Zrake, J.\ 2015,
Space Sci.\ Rev., 191, 77

\bibitem[Flower(1996)]{flow1996}
Flower, P.~J.\ 1996,
ApJ, 469, 355

\bibitem[Gonz\'alez et al.(2017)]{gonz2017}
Gonz\'alez, J.~F., Hubrig, S., Przybilla, N., et al.\ 2017,
MNRAS, 467, 437

\bibitem[Howarth \& Stevens(2014)]{how2014}
Howarth, I.~D., Stevens, I.~R.\ 2014,
MNRAS, 445, 2878

\bibitem[Hubrig et al.(2004a)]{Hubrig2004a}
Hubrig, S., Kurtz, D.~W., Bagnulo, S., et al.\ 2004a,
A\&A, 415, 661

\bibitem[Hubrig et al.(2004b)]{Hubrig2004b}
Hubrig, S., Szeifert, T., Sch{\"o}ller, M., et al.\ 2004b,
A\&A, 415, 685

\bibitem[Hubrig et al.(2008)]{Hubrig2008}
%
Hubrig, S., Sch\"oller, M., Schnerr, R.~S., et al.\ 2008,
A\&A, 490, 793

\bibitem[Hubrig(2013a)]{Hubrig2013a}
Hubrig, S.\ 2013a,
in: ``Massive Stars: From $\alpha$ to $\Omega$'', held 10--14 June 2013 in Rhodes, Greece;
online at http://a2omega-conference.net, id.~39

\bibitem[Hubrig et al.(2013b)]{Hubrig2013b}
Hubrig, S., Sch{\"o}ller, M., Ilyin, I., et al.\ 2013b,
A\&A, 551, A33

\bibitem[Hubrig et al.(2014a)]{Hubrig2014a}
Hubrig, S., Sch\"oller, M., Kholtygin, A.~F.\ 2014a,
MNRAS, 440, 1779

\bibitem[Hubrig et al.(2014b)]{Hubrig2014b}
Hubrig, S., Fossati, L., Carroll, T.~A., et al.\ 2014b,
A\&A, 564, L1

\bibitem[Leitherer \& Chavarria-K.(1987)]{leith1987}
Leitherer, C., Chavarria-K., C.\ 1987,
A\&A, 175, 208

\bibitem[Maheswaran \& Cassinelli(2009)]{mahes2009}
Maheswaran, M., Cassinelli, J.~P.\ 2009,
MNRAS, 394, 415

\bibitem[Mamajek(2008)]{mamajek2008}
Mamajek, E.~E.\ 2008,
AN, 329,10

\bibitem[Motte et al.(1998)]{motte1998}
Motte, F., Andre, P., Neri, R.\ 1998,
A\&A, 336, 150

\bibitem[Petit et al.(2013)]{petit2013}
Petit, V., Owocki, S.~P., Wade, G.~A., et al.\ 2013,
MNRAS, 429, 398

\bibitem[Pillitteri et al.(2014)]{pil2014}
Pillitteri, I., Wolk, S.~J., Goodman, A., Sciortino, S.\ 2014,
A\&A, 567, L4

\bibitem[Pillitteri et al.(2016)]{pil2016}
Pillitteri, I., Wolk, S.~J., Chen, H.~H., Goodman, A.\ 2016,
A\&A, 592, A88
%

\bibitem[Pillitteri et al.(2017)]{pil2017}
Pillitteri, I., Wolk, S.~J., Reale, F., Oskinova, L.\ 2017,
A\&A, 602, A92

\bibitem[Preston(1967)]{preston1967}
Preston, G.~W.\ 1967,
ApJ, 150, 547

\bibitem[Rho et al.(2004)]{rho2004}
Rho, J., Corcoran, M.~F., Hamaguchi, K., Lefloch, B.\ 2004,
ApJ, 607, 904

\bibitem[Rivinius et al.(2013)]{riv2013}
%
Rivinius, T., Townsend, R.~H.~D., Kochukhov, O., et al.\ 2013,
MNRAS, 429, 177

\bibitem[Sch\"oller et al.(2017)]{schol2017}
Sch\"oller, M., Hubrig, S., Fossati, L., et al.\ 2017,
A\&A, 599, A66

\bibitem[Skinner et al.(2008)]{skin2008}
%
Skinner, S.~L., Sokal, K.~R., Cohen, D.~H., et al.\ 2008,
ApJ, 683, 796

\bibitem[Snik et al.(2008)]{snik2008}
Snik, F., Jeffers, S., Keller, Ch., et al.\ 2008,
Proc.\ SPIE, 7014, E22

\bibitem[Stibbs(1950)]{stibbs1950}
Stibbs, D.~W.~N.\ 1950,
MNRAS, 110, 395

\bibitem[Townsend \& Owocki(2005)]{Townsend2005}
Townsend, R.~H.~D., Owocki, S.~P.\ 2005,
MNRAS, 357, 251

\bibitem[ud-Doula \& Owocki(2002)]{udDoula2002}
ud-Doula, A., Owocki, S.~P.\ 2002,
ApJ, 576, 413

\bibitem[ud-Doula et al.(2008)]{udDoula2008}
ud-Doula, A., Owocki, S.~P., Townsend, R.~H.~D.\ 2008,
MNRAS, 385, 97

\bibitem[van Leeuwen(2008)]{Leeuwen2008}
van Leeuwen, F.\ 2008,
VizieR Online Data Catalog: Hipparcos, the New Reduction, Cat.~1311
%

\bibitem[Vink et al.(2001)]{Vink2001}
Vink, J.~S., de Koter, A., Lamers, H.~J.~G.~L.~M.\ 2001,
A\&A, 369, 574

\bibitem[Wilking et al.(2008)]{wilking2008}
Wilking, B.~A., Gagn\'e, M., Allen, L.~E.\ 2008,
``Handbook of Star Forming Regions, Volume II: The Southern Sky'',
ASP Monograph Publications, Vol.~5. Edited by Bo~Reipurth, p.~351

%
%
%

%
%
%

%
%
%

%
%
%

%
%
%

\end{thebibliography}
\end{document}